\documentclass{elsart5p}

\newcommand{\eref}[1]{(\ref{#1})}

\newcommand{\bv}[1]{{\bf #1}}
\newcommand{\uv}[1]{{\bf \hat #1}}

\newcommand{\be}{\begin{equation}}
\newcommand{\ee}{\end{equation}}

\usepackage{graphics}
\usepackage{graphicx}
\usepackage{amssymb}
\begin{document}

\begin{frontmatter}



\title{Chirality induced anomalous-Hall effect in helical spin crystals}
%

\author[AA,BB]{B. Binz},
\author[AA,CC]{A. Vishwanath\corauthref{A. Vishwanath}},
\ead{ashvinv@socrates.berkeley.edu}

\address[AA]{Department of Physics, University of California, 366 Le Conte \# 7300, Berkeley, CA 94720 - 7300, USA}
\address[BB]{Institute of theoretical Physics, University of Cologne,
  Z\"ulpicher Str. 77, D-50937 Cologne, Germany}
\address[CC]{Material Sciences Division, Lawrence Berkeley
Laboratories, Berkeley CA}

\corauth[A. Vishwanath]{Corresponding author. Tel: (510) 643-3952
fax: (510) 643-8497}

\begin{abstract}
Under pressure, the itinerant helimagnet MnSi displays unusual
magnetic properties. We have previously discussed a BCC helical spin
crystal as a promising starting point for describing the high
pressure phenomenology. This state has topologically nontrivial
configurations of the magnetization field. Here we note the
consequences for magneto-transport that arise generally from such
spin textures.  In particular a skyrmion density induced
`topological' Hall effect, with unusual field dependence, is
described.

\end{abstract}


\end{frontmatter}


\section{Introduction}
Recently, the interplay between electron transport and magnetism has
been at the center of renewed activity. The anomalous hall effect in
ferromagnets is a prototypical example of this interplay.
Furthermore, transport in the presence of more complicated magnetic
order, such as spin spirals, are proving to be extremely rich. An
interesting analog of the anomalous Hall effect of ferromagnets, the
'topological' Hall effect, is expected \cite{AHE} where Berry's phase
from skyrmion spin textures leads to an effective magnetic flux.
Experimentally, novel physical phenomena have also been reported in
itinerant helimagnets (with spiral magnetic order), especially in
the B20 structure family which includes FeGe and MnSi. The latter
was long regarded as an example of a well understood magnetic
system, until recent experiments under moderate pressure revealed a
quantum phase transition into a 'partial order' state, with an
unusual signature in neutron scattering and non-Fermi liquid
resistivity signature $\Delta \rho(T) \propto T^{1.5}$
\cite{pfleiderer01,pfleiderer04}. This non-Fermi liquid resistivity
is seen for almost three decades in temperature from around 6K
down to a few mK. While the
unconventional magnetism disappears on further increasing the
pressure, the non-fermi liquid signatures in transport persists up
to the highest pressures studied. Similar phenomena are seen in the
related compound FeGe, where spin spirals are also believed to play
a role \cite{pedrazzini07}.


Recently, the anomalous Hall effect has been measured by several
groups both at ambient and higher pressures \cite{lee06}. M. Lee et
al. have identified a sharp feature in the Hall resistivity, which
appears to be increased in a  restricted parameter region of
temperature, pressure and field \cite{lee07}. They have put forward
the interesting speculation that this feature may be related to a
'topological' Hall effect (or Hall effect produced by spin
chirality) \cite{AHE}.

Here, we first review our proposal of BCC helical spin crystals as a
starting point to describing the `partial order'
state \cite{binz06,binz06b}. These and other spin crystals may be
understood as periodic arrangements of skyrmion configurations. The
physical manifestations of such topologically non-trivial spin
textures is the main subject of this article. In particular,   we
discuss the possibility of observing a topological Hall effect in
such textures, form two simple examples.

{\it Phenomenological theory of anomalous magnetism in MnSi at high
pressure:}  MnSi is found to have the following phase diagram. At
ambient pressure, a ferromagnetic ordering occurs below $30K$. The
presence of weak Dzyaloshinskii-Moriya interactions which are
allowed by the non-centrosymmetric B20 cubic structure, leads to a
spiraling of the magnetization. A single spiral state with a pitch
of about $\lambda \approx 170 \mbox{\AA}$ and a small moment
$0.3\mu_B$ is produced. Crystalline anisotropy locks the spiral
along certain fixed crystal directions - a simple symmetry analysis
reveals that the simplest energy terms consistent with the B20
crystalline symmetries are such that they favor the spiral along the
$\langle 111\rangle$  direction or the $\langle 100\rangle$
directions. Indeed, neutron scattering at ambient pressure reveals
Bragg spots along the $\langle 111\rangle$ direction (where $\langle
111\rangle$ refers to the [111] and the seven other
crystallographically equivalent directions). On increasing the
pressure beyond $p_c = 14.6kBar$, there is a quantum phase
transition into an unusual state termed the 'partial order' state in
Ref. \cite{pfleiderer04}. Here, in contrast to ambient pressure, the
elastic neutron scattering reveals enhanced scattering all around
the wavevector sphere $\bv{k} = 2\pi/\lambda$. However the
intensities are peaked around the $\langle 110\rangle$ directions.
It may be argued that this is very unlikely to be due to single
spiral states \cite{binz06,binz06b}.

We therefore adopt the next simplest hypothesis, that a useful
starting point to explain partial order is a spin crystal, made of a
coherent superposition of magnetic spirals along the $\langle
110\rangle$ directions.  Approaches which are complementary to ours
Refs. \cite{roessler06,fischer07}, have proposed different periodic
spin textures starting from real space rather momentum space. At
longer scales, magnetic order is believed to be destroyed by
disorder or fluctuation effects. Experimentally, while NMR sees a
frozen magnetic moment in this regime \cite{yu04}, $\mu$SR does not
\cite{uemura}. Here, we assume the order is static, at least in the
presence of a magnetic field.

Elsewhere, we have studied the energetics of such a state relative
to the single mode state \cite{binz06,binz06b}. Spatially varying
interactions between spirals are required to stabilize such a state,
and is more likely the shorter the spiral wavelength. In this
respect, although MnSi has a relatively long spiral wavevector, it
is the shortest compared to others in the same family (e.g. FeGe,
FeCoSi etc.). Here we do not revisit the energetics, but rather
adopt a phenomenological approach and deduce the consequences of
having a $\langle 110\rangle$  spin crystal. Note, the most general
such spin crystal can be specified by the amplitude and phase of the
spirals along each direction. This requires specifying six complex
numbers $\psi_1,\psi_2 \dots \psi_6$. From these, and a
specification of the basis used along each spiral direction, one can
reconstruct $\bv{M}(\bv r)$. If we choose the six wavevectors
$\bv{k}_j$ to be along
$\{(1,\bar{1},0),(\bar{1},\bar{1},0),(0,\bar{1},1),(0,\bar{1},\bar{1}),(1,0,1),(1,0,\bar{1})\}$,
and the basis vectors to be unit vectors $\uv\epsilon''_j \propto
[(0,0,1)\times \bv{k}_j]$, and $\uv\epsilon'_j \propto [\uv
\epsilon''_j \times \bv{k}_j] $, we can write:
\begin{equation}
 \bv{M}(\bv{r}) =\frac12 \sum_{j=1}^6 \psi_j (\epsilon'_j + i
\epsilon''_j)e^{i\bv{k}\cdot\bv{r}} + h.c.
\end{equation}
These spin structures have the periodicity of a body centered cubic
(bcc) lattice, hence, bcc spin crystals.

We can now develop a Landau theory in terms of these complex fields.
Cubic symmetry restricts the Landau free energy to the following
form: \be F=\sum_{jj'} V_{jj'}|\psi_j|^2|\psi_{j'}|^2 + \lambda\,
\mbox{Re} \left(T_x+T_y+T_z \right),\label{F} \ee where
$T_x=\psi_1^*\psi_2\psi_5\psi_6$, $T_y=\psi_1^*\psi_2^*\psi_3\psi_4$
and $T_z=-\psi_3\psi_4^*\psi_5^*\psi_6$. The bcc state is realized
if the parameters $V_{ij}$ and $\lambda$ are such that the energy
minimum has equal amplitudes for all six modes. As for the phases,
it is clear, that they only depend on the sign of the parameter
$\lambda$. For $\lambda>0$ $(<0)$, the phases arrange in such a way
that $T_x,T_y,T_z$ are negative (positive). The remaining three
phase degrees of freedom correspond to global translations of the
structure. Thus symmetry considerations alone lead to just two
possible bcc states, called bcc1 and bcc2 respectively. Both states
are doubly degenerate due to time reversal $\bv M\to-\bv M$. The
state bcc1 ($\lambda>0$) is more likely to be realized than bcc2,
because the magnitude of magnetic moments $|\bv M(\bv r)|$ has a
much narrower distribution in bcc1 than in bcc2 \cite{binz06}. The
resulting real-space pattern $\bv M(\bv r)$ of bcc1 is best
understood as a periodic network of skyrmion lines and anti-vortex
node-lines \cite{binz06,binz06b}. These structures are stabilized by
symmetry - the magnetic point group symmetry of the structure is
$O(T)$. This group is generated by $\pi/2$ rotations around the
$x$-, $y$- or $z$-axis, followed by
time-reversal \cite{binz06,binz06b}. We will see that the most direct
manifestation of these structures is in the magneto-transport,
leading to a  `topological' Hall effect.


\section{Skyrmion density}\label{skyrmion}

Non-trivial magnetic structures in metals can act as an effective
orbital magnetic field on conduction electrons and thus lead to an
anomalous Hall effect which is independent of spin-orbit coupling
\cite{AHE}. Essentially, when conduction electrons are forced to
follow the local spin direction $\bv M(\bv r)$, they acquire a
Berry's phase. This may be viewed as an effective field, which is
proportional to the skyrmion-density $\bv \Phi(\bv r)$, defined as
\be \Phi^\alpha=\frac1{8\pi}\epsilon^{\alpha\beta\gamma}\,\uv
n\cdot\left(\partial_\beta\uv n\times\partial_\gamma\uv
n\right),\label{Phi} \ee where $\uv n=\bv M/|\bv M|$. The flux of
$\bv \Phi$ through any two-dimensional rectangle with periodic
boundary conditions is  a topological winding number, i.e. an
integer. Note, the terminology 'spin chirality' is somewhat
misleading since $\bv \Phi$ transforms as an axial vector, like the
magnetic field. It is {\it even} under inversion, hence not a chiral
object - i.e. broken inversion symmetry is not required to have $\bv
\Phi\neq 0$.

It is remarkable that so many different theoretical proposals for
novel magnetic structures in 3D magnets without inversion symmetry
independently lead to arrangements of skyrmion lines (meron lines,
double-twist cylinders) \cite{binz06,binz06b,roessler06,fischer07}.
Therefore, such systems appear to be natural candidates for
observing the effects of  $\bv \Phi$.

By inspection of Eq. \eref{Phi}, it becomes immediately clear that a
simple spin density wave, which involves a single pair of
wavevectors $\pm \bv k$, always has a {\it vanishing} skyrmion
density, since in this case $\bv M$ depends only on a single space
coordinate. This is also true if $\bv M$ acquires a uniform
component, i.e. in the conical phase under a magnetic field. Multi-q
spin crystals are therefore the simplest slowly varying magnetic
structures with non-vanishing skyrmion density.

\subsection{Square spin crystal}

As the simplest example for helical spin crystal, we first consider
the magnetic texture obtained by superimposing two orthogonal
spirals of the same chirality. For simplicity, we choose units such
that each spiral has unit amplitude and a wave-length equal to
$2\pi$. In appropriate coordinates, the square spin crystal is then
given by \begin{equation} \bv M(\bv r)=(\sin y, \sin x, \cos y -
\cos x) \label{square_crystal}\end{equation}

\begin{figure}
\begin{center}
\includegraphics[scale=0.6]{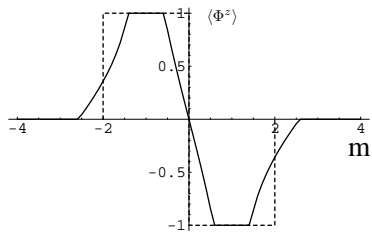}
\caption{The dotted line shows the skyrmion number   $\int \!dxdy
  \Phi^z$ of one unit cell for the helical 2D
  square spin crystal with a uniform component $\bv
M(\bv r)=(\sin y, \sin x, m+\cos y - \cos x)$  as a function of $m$.
The full line shows the same quantity evaluated with a cutoff for
$\bv M^2<1/3$, as explained in the text. \label{sqPhi}}
\end{center}
\end{figure}

 Note, it is a 2D like structure given that there is no variation along the
$z$ direction. This is a periodic 2D arrangement of skyrmion-like
textures pointing up and down  in a checkerboard pattern. The
skyrmion density $\Phi^z$ follows this pattern, in fact,
$\Phi^z=M^z/(4\pi |\bv M|^3)$ in this case. Thus for symmetry
reasons, the total skyrmion number over the unit cell ($\int
\!dxdy\, \Phi^z$) vanishes, since positive and negative
contributions cancel exactly. Now, a magnetic field is applied in
the z-direction. We assume, in the interest of clarity, that this
only adds a uniform magnetization $(0,0,m)$ to $\bv M$, but does not
otherwise affect the spin arrangement. The skyrmion number per unit
cell is an integer, and therefore can not change continuously with
$m$. The behavior is shown in Fig. \ref{sqPhi}. The function is
discontinuous at $m=0,\pm2$ since it is at precisely these values of
$m$ that the spin texture develops nodes. These values can be easily
obtained from the  $m=0$ magnetization pattern of Eq.
\ref{square_crystal} by noting that the magnetization that is purely
along the $z$ axis takes on the values $\bv M =0,\pm 2\uv{z}$. In
the parameter regime $0<|m|<2$, a non-trivial, quantized spin
chirality is obtained, $\Phi^z=-{\rm sgn}(m)$.

At least in the limit of strong coupling between conduction
electrons and magnetic background, it is natural that the spin
direction $\uv n$ rather than the amplitude enters  Eq. \eref{Phi}.
However, in contrast to Refs. \cite{AHE}, we are concerned with a
smoothly varying magnetization $\bv M(\bv r)$, which has a spatially
varying amplitude and even nodes. In the vicinity of nodes,
electrons can not be infinitely sensitive to directional changes of
the magnetic background and the theory needs to be extended. To
include this effect in a simple way, let us assume that the
electrons couple only then to the background spin texture, if the
latter has an amplitude bigger than some cutoff. I.e. we consider
$\bv \Phi_c=\Theta\left(\bv M^2-M_0^2\right)\bv \Phi$, where
$\Theta$ is the Heaviside step function. As a result of this
modification, the discontinuous jumps of the skyrmion number gets
smoothed out, as shown in Fig. \ref{sqPhi}.

\subsection{bcc1 spin crystal}

We now consider the spin crystal bcc1. An explicit expression for
$\bv M(\bv r)$ is given in Ref. \cite{binz06b}. Again, the average
of the skyrmion density over a unit cell is zero for symmetry
reasons [the point group symmetry of bcc1 is $O(T)$ and therefore
does not tolerate a nonzero axial vector like $\langle\bv
\Phi\rangle$]. Hence, there are local skyrmion densities and
conduction electrons will feel their effective fields, but the
average effective field is zero. No anomalous Hall effect can
therefore occur without an applied field.

Applying an external magnetic field lowers the symmetry and allows
for $\langle\bv \Phi\rangle\neq0$. An external field has two effects
on the bcc1 structure. First, it will induce a uniform magnetization
$\bv m$  pointing along the field and reduce the amplitude of the
helical components. Second it will affect the relative amplitude and
phases of the six interfering helical modes, thus transforming the
spin texture. Eventually, the magnetic field will induce a
transition to a different magnetic state, for example the conical
single-spiral state or the spin-polarized (ferromagnetic) state.
Here, we will only discuss the effect of a uniform component  $\bv
m$ added to the otherwise unperturbed bcc1 structure.

Before we go into the details, we make a short remark about time
reversal symmetry (T). The bcc1 state breaks T in such a way that it
may not be restored by a subsequent translation
\cite{binz06,binz06b,binz07}. As a consequence, there are two
degenerate bcc1 states related by time reversal: $\bv M( \bv
r)\to-\bv M(\bv r) $. Let us label them by a parameter $S=\pm1$.
Time reversal symmetry requires that the integrated skyrmion density
of the state $S$ plus a uniform component $\bv m$ satisfies
 \be
\left. \langle\bv \Phi\rangle\right|_{S,\bv m}=- \left. \langle\bv
\Phi\rangle\right|_{- S,- \bv m}\label{T}
 \ee

 In the following, we choose the lattice constant of the cubic
(i.e. not elementary) unit cell as the unit of length and the
amplitude of the elementary spiral modes which build up bcc1 as the
unit of magnetization. The net effective field is determined by the
integral of $\bv \Phi$ over the unit cell. It may be trivially
written as \be \langle\Phi^z\rangle=\int\!dz\, n_z(z) \label{nz}\ee
where $n_z=\int\!dxdy\,\bv \Phi$ is an integer-valued function of
$z$. It can only change its value at those positions of $z$, which
correspond to nodes of $\bv M$. For $\bv m=0$,  the  bcc1 structure
 has node lines, and $n_z(z)$ could potentially change
everywhere. Fortunately, we know from symmetry that in this case
$\langle\bv \Phi\rangle=0$. For $\bv m\neq0$, there are only
discrete point nodes, and Eq.  \eref{nz} turns out to be very useful
to evaluate $\langle\Phi^z\rangle$. The other components,
$\langle\Phi^{x,y}\rangle$  may be obtained in a similar way.

\begin{figure}
\begin{center}
\includegraphics[scale=0.6]{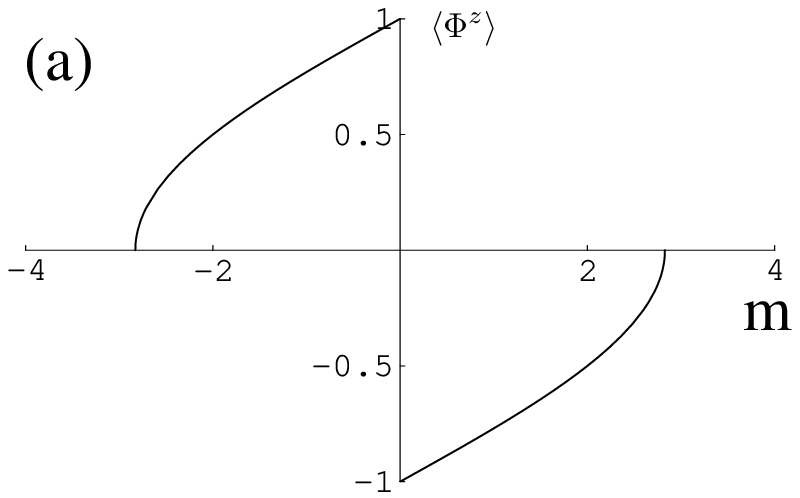}
\includegraphics[scale=0.6]{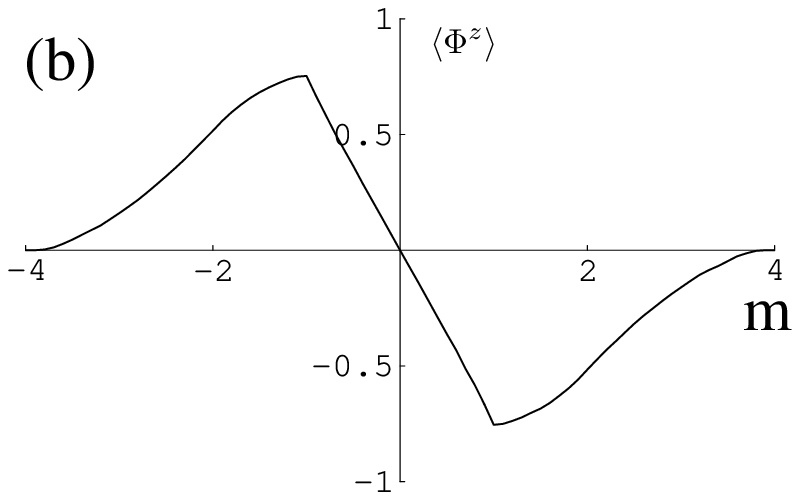}
\caption{Spin Chirality for bcc1 structure with a uniform
magnetization $(0,0,m)$ superimposed (a) Shows
$\langle\Phi^z\rangle$ as function of $m$. (b) Same with a
  cutoff for $\bv M^2<1$.
\label{bccPhi}}
\end{center}
\end{figure}

As an example, we consider the case $\bv m=(0,0,m)$, i.e. a uniform
component along $\uv z$ is added to the bcc1 texture (the case with
the field along (111) direction is found to be qualitatively
similar). For this geometry, $\langle\bv \Phi\rangle$ is an odd
function of $m$ and parallel to $\uv z$, because of the $O(T)$ point
group symmetry. Hence, by Eq. \eref{T}, the result is independent of
the time-reversal label $S$. The nodes of $\bv M$ are particularly
easy to analyze as a function of $m$. As a result,
$\langle\Phi^z\rangle$ may be calculated analytically. We find
$\langle\Phi^z\rangle=-{\rm sgn}(m)\,(2/\pi)\arccos(|m|/2/\sqrt{2})$
for $0<|m|<2\sqrt{2}$ and $\langle\Phi^z\rangle=0$ elsewhere. The
behavior is shown in Fig. \ref{bccPhi}. For comparison, the maximum
spin density $|\bv M|$ at $m=0$ is $2\sqrt{3}$. There are two phase
transitions as a function of the uniform component. One is of course
at $m=0$, where $\langle\Phi^z\rangle$ changes sign. However, the
nontrivial transition occurs at the high-field ends, where
$\langle\Phi^z\rangle$ vanishes continuously but with an infinite
slope. Note, at this point the magnetic structure is still
non-trivial, but the 'spin-chirality' has been ironed out by the
large uniform magnetization.
 Like in the former example,
introducing a cutoff magnitude $M_0$, to model the decoupling of
electrons from the low magnetization parts of the textures, smoothes
these transitions to some degree. For the BCC1 spin crystal, with a
spiral wavelength of 170{\AA}, the 'spin-chirality' induced
effective field is about $1.7$Tesla.

\section{Summary}

We have investigated the possibility of a topological Hall effect in
non-trivial magnetic structures composed of skyrmion configurations
under the influence of an external magnetic field. In two simple
examples, we have analyzed the integrated skyrmion density
$\langle\bv \Phi\rangle$, which has been shown to act on conduction
electrons as an effective internal field.

In both cases it is clear that an induced magnetization plays two
opposite roles. On the one hand, it is required by symmetry to
induce a non-zero spin chirality, but on the other hand, increasing
it beyond a threshold value leads to a vanishing $\langle\bv
\Phi\rangle$. This will ultimately lead to a non-monotonic effect of
magnetization on the Hall conductivity, which is influenced by
$\langle\bv \Phi\rangle$. This, we believe is a hallmark of the spin
chirality contribution to the Hall effect.

In the case of a two-dimensional spin structures, which may be
realized for example in thin films, magnetic surfaces or interfaces,
There is the possibility of a plateau region, where the effect is
independent of the external field, due to quantization of the
skyrmion number.

In both above cases, the effective field created by spin
configurations turns out to be antiparallel to the uniform
magnetization. This  predicts the topological Hall effect to be of
opposite sign to the normal Hall effect.

Questions left for future investigation include whether the
`topological' Hall effect discussed here is relevant to the
experiments of \cite{lee07}, which are still on the low pressure
side, and how electron transport is affected by a disordered, and
possibly fluctuating, arrangement of non-trivial spin textures.
\section{Acknowledgement}
We would like to thank Y. B. Kim and Minhyea Lee for useful
discussions and especially Phuan Ong for sharing his insights into
the topological hall effect in metals. This work is supported in
part by the Hellman Family Faculty fund, A. P. Sloan Foundation
fellowship and LBNL DOE-504108 (AV), and the Swiss National Science
Foundation (BB).

\end{document}